# Modeling materials with optimized transport properties

**Peter Kratzer,[1] Vladimir M. Fomin,[1,2] B. Hülsen[3] and M. Scheffler[3]**

**[1]Faculty of Physics, University Duisburg-Essen, Lotharstr. 1, 47048 Duisburg, Germany**
**e-mail: Peter.Kratzer@uni-duisburg-essen.de**
**[2]Institute for Integrative Nanosciences, IFW Dresden, 01069 Dresden, Germany**
**[3]Fritz-Haber-Institut der Max-Planck-Gesellschaft, 14195 Berlin, Germany**

## ABSTRACT

Following demands for materials with peculiar transport properties, e.g. in magnetoelectronics or thermoelectrics, there is a need for materials modeling at the quantum-mechanical level. We combine density-functional with various scale-bridging tools to establish correlations between the macroscopic properties and the atomic structure of materials. For examples, magnetic memory devices exploiting the tunneling magneto-resistance (TMR) effect depend crucially on the spin polarization of the electrodes. Heusler alloys, e.g. $Co_2MnSi$, if perfectly ordered, are ferromagnetic half-metals with (ideally) 100% spin polarization. Their performance as electrodes in TMR devices is limited by atomic disorder and deviations from perfect stoichiometry, but also by interface states at the tunneling barrier. We use *ab initio* thermodynamics in conjunction with the cluster expansion technique to show that excess manganese in the alloy and at the interface helps to preserve the desired half-metallic property. As another example, nanostructured materials with a reduced thermal conductivity but good electrical conductivity are sought for applications in thermoelectrics. Semiconductor heterostructures with a regular arrangement of nanoscale inclusions ('quantum dot superlattices') hold the promise of a high thermoelectric figure of merit. Our theoretical analysis reveals that an increased figure of merit is to be expected if the quantum dot size, the superlattice period and the doping level are all suitably fine-tuned. Such a superlattice thus constitutes a material whose transport properties are controlled by geometrical features at the nanoscale.

## 1. Introduction

Novel materials, in particular alloys or nanostructured materials, or heterostructures combining several materials into one, offer ample opportunities to find innovative solutions for devices with peculiar, tailored electrical transport properties. However, optimizing these materials or structures with respect to a particular figure of merit still poses a challenge to simulations: Because of the heterogeneity inherent in disordered alloys or in nanostructured samples, large systems, with possibly up to a hundreds of thousands of atoms, need to be modeled, and/or a huge configuration space needs to be explored. Moreover, it is crucial to consider quantum mechanics for electronic transport properties: Electronic states different from those known from the homogeneous bulk materials may arise due to the modification of the electronic band structure both by the local chemical environment in alloys and by the quantum confinement in nanostructures, or due to the quantum nature of electron tunneling through a potential barrier. Thus we need to use simulation tools that make use of information on the atomic and electronic scale, while at the same time taking into account varying materials properties, such as strain and composition fluctuations, on a much larger scale. In this paper, we will present examples of hierarchical multi-scale modeling that enables us to combine information about the systems under study from different length scales, ranging from quantum effects on band structure and energy levels to macroscopic transport properties.

## 2. Formation energies, electronic and magnetic properties from a cluster expansion

Many systems relevant for electronic transport are crystalline in nature, i.e. the atoms are located on well-defined lattice positions, while displaying spatial inhomogeneities originating from various chemical species occupying the lattice sites, or from coherent nanoscale inclusions in an otherwise homogeneous host material. Retaining crystallinity is often crucial for achieving good electrical conductivity, because trapping of carriers or scattering due to defects usually severely limits the achievable conductivity in samples that are amorphous or have a poor crystallinity. In some of these crystalline systems, mechanical strain due to inhomogeneities is an important issue affecting the electronic structure, for example in semiconductor heterostructures with a lattice mismatch of the constituents. A possible way to treat these strain effects will be described further below. First, we will consider alloys where the local chemical environment of the constituent species is the dominant factor determining their electronic properties. Here, the cluster expansion technique can be helpful in tackling the complexity introduced by the huge number of possible atomic configurations that results from occupying each of the lattice sites randomly by one of the various species.

To give a specific example, we consider the Heusler alloy $Co_2MnSi$ and its variants obtained by varying the relative concentrations of Co and Mn in this alloy. This material is of special interest, since density-functional theory calculations [1] predict the ideal $Co_2MnSi$ alloy with the $L2_1$ crystal structure to be a ferromagnetic half metal. This means that this material has electronic states at the Fermi level only in one spin channel, while displaying a band gap in the other spin channel. This makes $Co_2MnSi$ attractive as material for spintronics, for example as electrode material in magnetic random-access memories that use the tunneling magneto-resistance (TMR) effect for the electronic read-out of the stored information. We are posing the question if the property of half-metallicity is stable when antisite defects, for example due to thermal disorder, are present, or when the relative concentration of Mn and Co in the alloy is varied.

The cluster expansion (CE) is a mathematical tool [2] that expresses a property of a system defined on a lattice by an expansion into an (in principle infinite) series of multi-site interactions, such as singles, pairs, triples, etc. This technique is useful, as one expects that, due to the 'near-sightedness' of nature, an expansion into a finite number of terms (a finite number of 'figures', as the building blocks of the expansion are called) will be sufficiently accurate for many systems and properties of interest. However, one should keep in mind that the validity of the series truncation needs to be checked carefully for each system and property one wants to investigate.

Since excellent reviews of the CE technique can be found in the literature [3], we only briefly explain the basic idea. Let us consider a binary alloy $A_{1-x}B_x$ with $N$ lattice sites. One particular configuration of the whole crystal is described by the occupation vector $\boldsymbol{\sigma} = \{\sigma_1, \sigma_2, \ldots, \sigma_N\}$, where $\sigma$ is +1 (–1) if a lattice point is occupied by atom A (B). For some property $F$ of interest, e.g. the formation energy, the magnetic moment, etc., we may construct an expansion similar in structure to the Hamiltonian of the Ising model,

$$F(\sigma) = J_0 + \sum_i J_i \sigma_i + \sum_{i,j} J_{ij} \sigma_i \sigma_j + \sum_{i,j,k} J_{ijk} \sigma_i \sigma_j \sigma_k + \ldots,$$

where the pairs, triplets and higher order terms are the `figures' of the CE. The effective cluster interactions $J_i$ of the CE are determined from the results of a relatively small number of configurations (≈50) obtained through first-principles computations. This can be achieved by a least-square fit of the predicted and the calculated values of $F(\sigma)$ on subsets of the structures calculated within DFT [4]. Since in practice one is working with a truncated expansion with a finite number of `figures', it is essential to check the transferability of the fit to other (unknown) configurations. A quantitative indicator of transferability is the cross-validation score, i.e. the average mean-square deviations on subsets for which DFT data are available, but have not been included in the fit in the first place. The set of figures that minimizes the cross-validation score is termed the optimal cluster expansion.

Once an optimum cluster expansion has been set up, the CE is a computationally inexpensive tool to explore the properties for a large number of configurations. A frequently studied property is the formation energy of alloys. There, the cluster expansion is useful in identifying ordered intermetallic compounds with a large supercell. In contrast to other computationally cheap simulation tools suitable to a large number of atoms, e.g. molecular dynamics using classical interatomic potentials, the cluster expansion may be used to represent also properties of quantum-mechanical origin directly in a simple and efficient way. For instance, a cluster expansion has been employed to search for configurations with high Curie temperature in the dilute magnetic semiconductor Ga(Mn)As [5]. Previously, this technique has also been used for finding semiconductor heterostructures with a prescribed band gap for given chemical constituents of the material [6]. More recently, we investigated the role of composition fluctuations in $CuIn_xGa_{1-x}Se_2$ for the local variations of the band gap in solar cells [7] using the CE method.

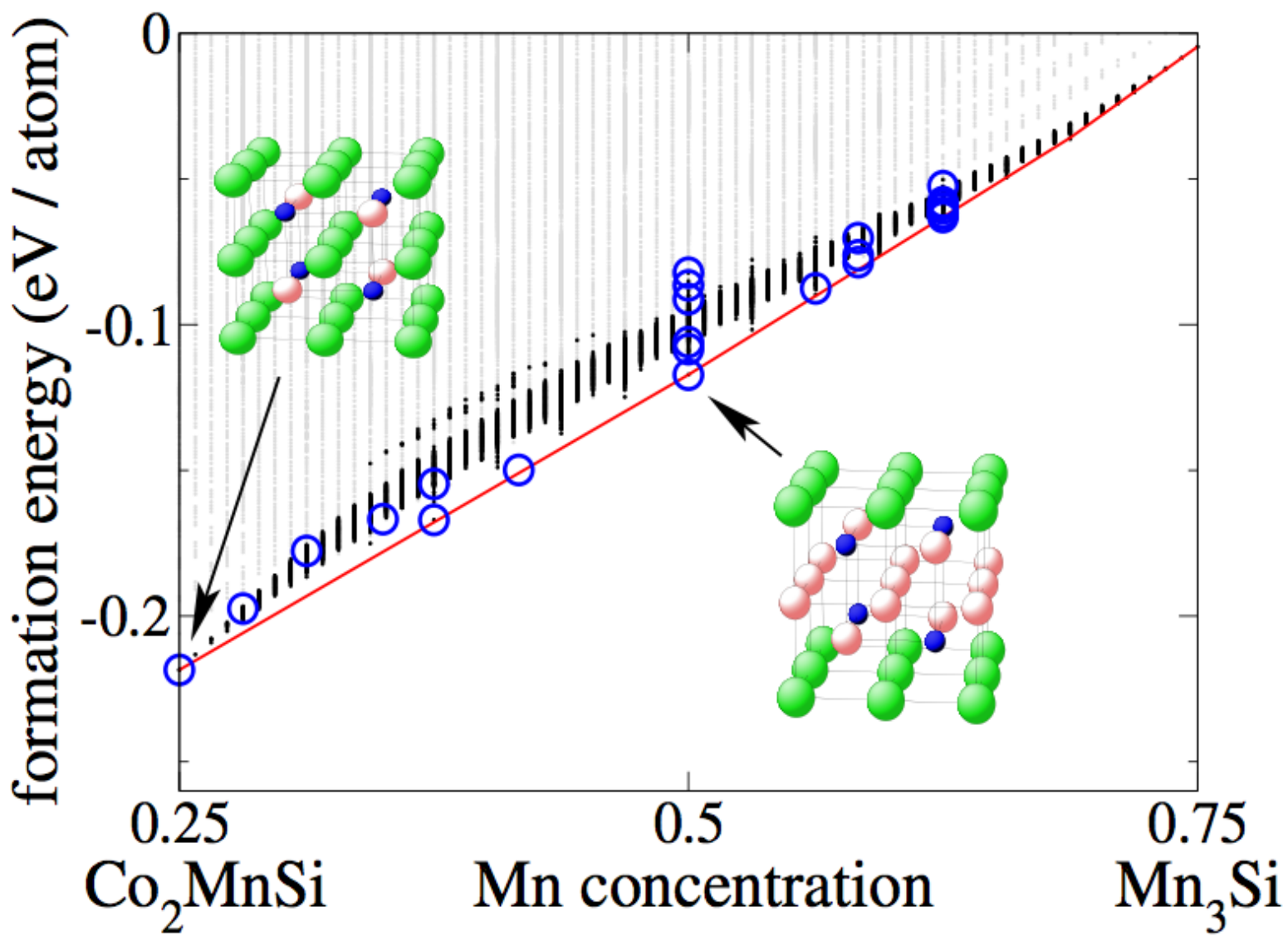

Figure 1. Formation energies of 27 million structures obtained from the cluster expansion by direct enumeration (light grey points). For the black points, the cluster expansion of the magnetic moments predicts an integer value. Both the integer magnetic moment and the existence of a spin gap has been confirmed by DFT calculations for the structures indicated by the circles.

In our present case of the Heusler alloy $Co_2MnSi$, we use a cluster expansion of the magnetic moment of the supercell [8]. Since a magnetic half-metal has an integer magnetic moment (an integer number of electrons in the minority spin channel that displays the energy gap), the magnetic moment serves as a necessary (although not sufficient) indicator of alloy compositions that possibly show half-metallicity. We find that the exchange of Co by Mn atoms in the alloy most likely preserves the half-metallic property in a range of compositions. This is concluded from the results shown in Fig. 1, where the light gray data points indicate the formation energies of 27 million different configurations screened by the cluster expansion, while the black data points indicate those configurations for which the cluster expansion predicts an integer magnetic moment. It is remarkable that the region of black points has a certain width on the energy scale, indicating that even a configuration different from the ground state, that could result from a thermally excited atomic configuration or from imperfect annealing of the sample, might still preserve the half-metallicity for some Mn-enriched Heusler alloys. Since the integer magnetic moment is just a first indicator of half-metallicity, additional DFT calculations have been performed to test some selected configurations (circles in Fig. 1). Indeed an energy gap in the Kohn-Sham eigenvalue spectra of the minority spin channel has been found in all cases. In summary, Fig. 1 establishes that adding more Mn in the synthesis of $Co_{2-x}Mn_{1+x}Si$ may even enhance its half-metallic properties. The robustness of Mn-enriched $Co_2MnSi$ alloys for spintronics applications has been confirmed in a recent experimental study [9].

## 4. Tunneling conductivity

One important application of half-metallic Heusler compounds is in TMR elements where two magnetic electrodes are separated by a very thin (only a few nanometer thick) oxide barrier. The relative magnetization of both electrodes is used to represent the bit of information stored in the TMR element, while the electrical conductivity of the tunnel junction serves as the read-out signal. The figure of merit is the TMR ratio, i.e. the difference of the conductance through the TMR device in the two states where the electrodes are magnetized either parallel or antiparallel to each other, divided by the smaller of the two values. Controlling the half-metallicity in the bulk of the electrodes by suitable alloy composition and materials processing (deposition and tempering steps) is prerequisite for achieving a high figure of merit. However, electronic states inside the gap in the minority spin channel at the interface between electrode and oxide barrier could act as centers for spin-flip scattering and could thus significantly diminish the figure of merit. In order to control and possibly eliminate this detrimental effect, one needs to know whether these interface states are localized at only one side of the oxide barrier (in the case of antiparallel magnetization of the electrodes), or if they extend through the barrier, and how much they contribute to the transmission of electrical current. The electronic properties of the interface states depend on their energetic position relative to the Fermi energy in the electrodes and on their orbital symmetry. The latter aspect is important for epitaxial, highly crystalline barriers made of MgO: An *s*-like character of the interface state allows for hybridization between the transition metal orbitals and the states derived from the conduction band in MgO, leading to metal-induced gap states, while a 3*d*-like character of these states prevents their hybridization with any states near the fundamental band gap in the MgO. Which of these alternatives is realized is a question that can be answered by DFT calculations of the interface electronic structure.

Of course, the results of such calculations depend on the atomic structure of the interface. First, the energetically most favorable interface structure needs to be determined. Depending on the conditions under which the Heusler electrodes are prepared, i.e., if there is a surplus of Co or Mn, we find that the alloy will be terminated at the interface either by a Co layer (Co-rich conditions) or by a mixed MnSi layer (Co-poor conditions) [10]. Calculation of the Kohn-Sham band structure shows that electronic interface states in the minority spin channel occur for both of these terminations. An example is shown in Fig. 2 for the interface layer of Co atoms bonding to the oxygen atoms in the top-most MgO layer. However, the electrons in this state must tunnel through the MgO barrier with a finite crystal momentum parallel to the interface, because the interface band crosses the Fermi level amid the ΓM and ΓX lines in the Brillouin zone (see Fig. 2, left panel). Moreover, the interface state has mostly Co-3*d*-orbital character. For both reasons, the wavefunction amplitude of the interface state is suppressed inside the MgO barrier (Fig. 2, right panel). Only if the $Co_2MnSi$ electrode is terminated by a full Mn layer, which is not stable in thermodynamic equilibrium, we find that the gap remains free of interface states.

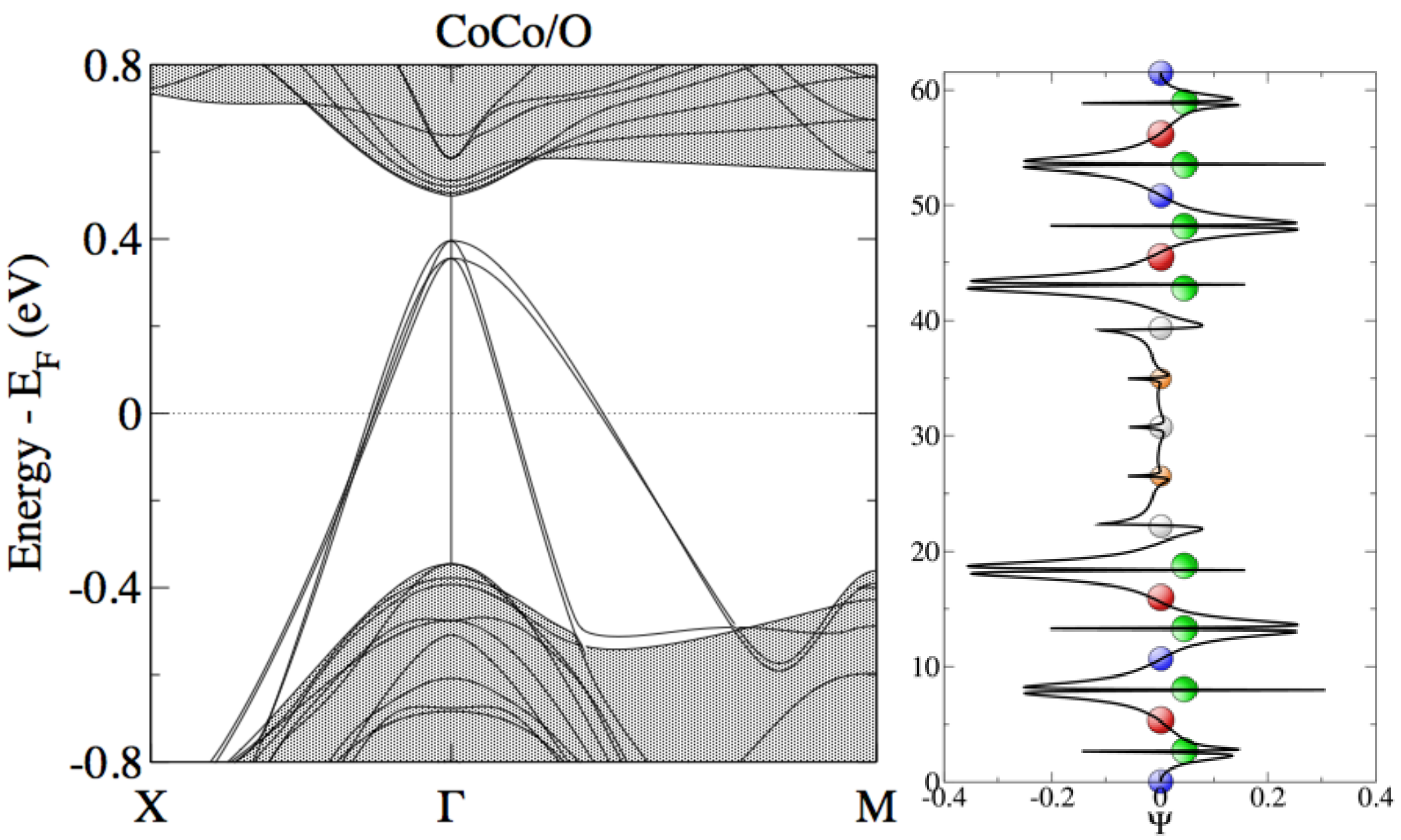

Figure 2. Kohn-Sham band structure projected onto the Brillouin zone of the interface in a $Co_2MnSi(001)/MgO(001)/Co_2MnSi(001)$ heterostructure terminated by Co layers. The gap region in the minority spin channel is shown. The shaded regions correspond to projected bulk bands of $Co_2MnSi$. The highly dispersive bands inside the gap are Co-induced interface states. The right panel shows the wavefunction belonging to the interface band at $E_F$. The vertical axis is a spatial coordinate normal of the interface. The small balls in the middle of the picture symbolize the Mg and O atoms of the barrier.

**4. Miniband transport in nanostructured materials**

While tunneling conductance may be considered a quite special case, understanding the conductivity of a material in general not only requires knowledge about its electronic structure, but also about the mechanisms of energy and momentum relaxation, e.g. scattering of carriers by phonons and/or impurities. A truly first-principles modeling of the electric transport properties of real materials is therefore a very challenging task, and presently we can only present some steps towards a full theoretical treatment, with many steps left to future work. The first and mandatory step is always a quantum-mechanical modeling of the electronic structure, since ordered alloys or heterostructures may display electronic states substantially different from those of pure bulk materials, thus providing new channels for electronic transport.

Here, we present an example where nanostructuring a material has a pronounced effect on both its electronic structure and the mechanisms available for scattering of carriers: Semiconductors with a regular array of nanoscale inclusions coherent with the host lattice ('epitaxially self-assembled quantum dot crystals' [11]) display one-dimensional electronic minibands [12] that act as additional channels of conductivity. At the same time, we need to take into account that the electron-phonon scattering in these minibands is substantially different from scattering in the bulk conduction band, mostly due the different phase space available for scattering in either case. In order to treat this complex, nanostructured system, we employ a multi-step approach [13]: First, atomistic modeling is used to describe the mechanical strain resulting from the nanoscale inclusions in the sample. The effect of the nanoscale confinement of the electrons in the quantum dots (QDs) and of the strain in both the quantum dots and the host matrix on the electronic structure is calculated quantum-mechanically from a tight-binding Hamiltonian. Thus, both the energetic position and the dispersion of the minibands are taken from a microscopic theory. On another level of modeling, for the calculation of the electron-phonon-scattering rate, we use a theory that is still state-specific, i.e. it calculates the scattering rate between quantum states with specific miniband index and crystal momentum, but uses a more coarse-grained input for evaluating the scattering matrix elements: The phonon dispersion is taken from an acoustic effective-medium theory, and the electronic wavefunctions are adopted from the analytic solution of a Kronig-Penney model of particles in a periodic array of quantum wells. The numerical value of the electron-phonon coupling strength is taken from experimental data. While all these pieces of information could in principle be obtained from the microscopic model as well, incorporating some level of empiricism allows us to perform the calculations faster and to explore systematically a variety of nanostructures differing in their geometrical parameters.
Specifically, the simulations have been carried out for atomistic models of the structures shown schematically in Fig. 3. One supercell consists of about 50,000 atoms. The relaxation of the atomic positions is performed using an Abell-Tersoff-type force field [14]. The orthogonal tight-binding Hamiltonian builds on an $sp^3s^*$ representation of the hopping matrix elements including first and second neighbors [15]. Spin-orbit coupling is included on an empirical level. The power-law dependence of the hopping matrix elements on interatomic distances [16] allows us to incorporate the effect of strain on the electronic states directly by setting up the Hamiltonian using the relaxed atomic positions. Periodic boundary conditions are employed to calculate Bloch states of the QD crystal. The folded-spectrum method is used to extract single eigenstates energetically located inside the band gap of the GaAs host, which physically correspond to the minibands in the QD stack.

## 4. Nanostructured thermoelectrics

For applications in thermoelectric converters, one is interested in materials that maximize the dimensionless figure of merit *ZT*. It describes the maximum available electrical power for a given temperature gradient across the converter, and is defined as

$$ZT = \frac{\sigma S^2}{\kappa_{\mathrm{el}} + \kappa_{\mathrm{ph}}} T.$$

Here, $\sigma$ is the electrical conductivity, $S$ is the Seebeck coefficient, $\kappa = \kappa_{\mathrm{el}} + \kappa_{\mathrm{ph}}$ is the thermal conductivity, comprised of an electronic contribution and a lattice contribution, and $T$ is the absolute temperature. Optimally suited materials for thermoelectric applications should have a low thermal conductivity, but simultaneously a high electrical conductivity. These are requirements that are difficult to meet in bulk materials, where in many cases the electrical and the thermal conductivity (the electronic contribution thereof) are interrelated, e.g. by the Wiedemann-Franz law in metals. Obviously it is difficult to find a general strategy for optimizing *ZT*, but previous work suggests that materials with a narrow electronic band at the Fermi energy could allow one to achieve a high *ZT* value, both due to a high value of $S$ in a narrow band, and a decoupling of $\sigma$ and $\kappa$ when scattering of carriers is only possible within a bandwidth smaller than $k_B T$ [17,18]. Arrays of self-assembled semiconducting quantum dots embedded in a semiconductor matrix are relevant in this context for several reasons: First, the nanoscale inclusions may act as very efficient scatterers of phonons with long and medium wavelength, thus suppressing thermal conductivity. For example, control of thermal conductivity down to < 1W/(m K) range at the nanoscale via individual phonon scattering barriers has been recently achieved in multilayered Ge/Si QD arrays with as few as five barriers [19]. Secondly, one-dimensional electronic minibands for the motion perpendicular to the growth direction are formed if the spacer layers between subsequently grown layers of QDs are very thin, less than about ten nanometers. These narrow minibands open up the possibility to optimize the figure of merit [20], following the strategy outlined above.

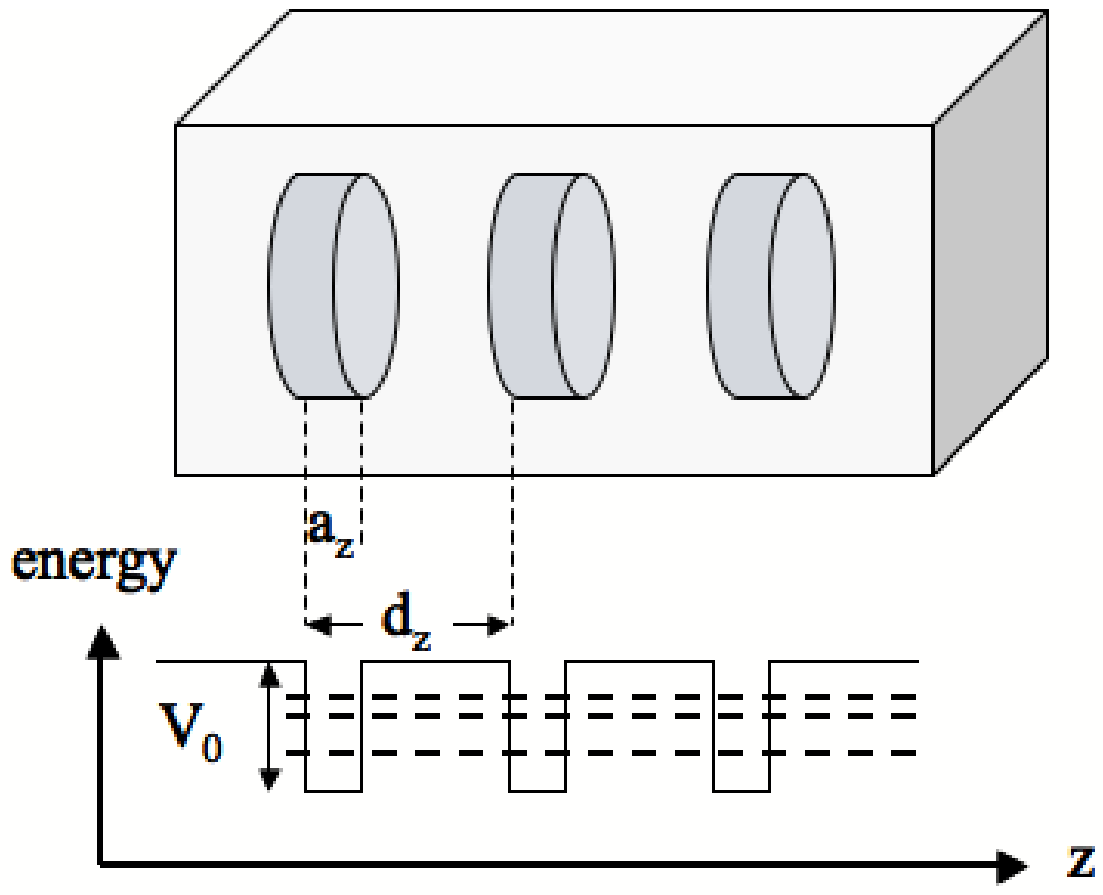

Figure 3. Schematic drawing of a stack of quantum dots of height $a_z$, with stacking period $d_z$ (upper part). The sample was grown along the $z$-axis. The lower part schematically shows the periodic confining potential and the energetic position of the minibands (dashed horizontal lines).

In the following, we describe our search for optimal structural parameters of the QDs for the case of InAs QDs embedded in GaAs. We consider a one-dimensional stacking sequence, with period $d_z$, of QDs with height $a_z$ (cf. Fig. 3). The parameter space of $a_z$ and $d_z$ has been explored in an attempt to optimize the figure of merit $ZT$, while the cylindrical shape of the QDs with a diameter of 10 nm has been kept for the sake of simplicity. As conceptual framework for calculating the transport coefficients $\sigma$, $S$ and $\kappa_{\mathrm{el}}$ entering $ZT$, we employ the semiclassical description of transport by the Boltzmann equation. Within each miniband, an energy-dependent carrier relaxation time $\tau_J$ is used in solving Boltzmann's equation to first order in both the electric field $E$ and the temperature gradient $\nabla T$ by linear expansion around the equilibrium carrier distribution $f_0$ :

$$v_{z,J}(k_z)\left[eE + \frac{\varepsilon_J(k_z) - \mu}{T}\nabla T\right]\left(-\frac{\partial f_0}{\partial \varepsilon}\right) = \left(-\frac{\partial f_0}{\partial \varepsilon}\right)_{\mathrm{coll}} \approx -\frac{f - f_0}{\tau_J(k_z)}.$$

We find that the carrier relaxation time, as calculated from Fermi's Golden Rule for the electron-acoustic-phonon scattering, displays significant structure inside the energy range of each miniband. Consequently, the approximation of a constant, energy-independent relaxation time, which is often made in calculating bulk conductivity, cannot be used for conductivity due to minibands. Within Boltzmann's theory, moments of the differential distribution function can be calculated,

$$L^{(\alpha)} = \sum_J \frac{2}{A} \int_0^{\pi/d_z} \frac{dk_z}{2\pi} \left( -\frac{\partial f_0}{\partial \varepsilon} \right) \tau_J(k_z) v_{z,J}^2(k_z) (\varepsilon_J(k_z) - \mu)^\alpha$$

and $\sigma$, $S$ and $\kappa_{\mathrm{el}}$ can be expressed in terms of these moments:

$$\sigma = e^2 L^{(0)},$$
$$S = -\frac{1}{eT}\frac{L^{(1)}}{L^{(0)}},$$
$$\kappa_{\mathrm{el}} = \frac{1}{T}\left( L^{(2)} - \frac{\left(L^{(1)}\right)^2}{L^{(0)}} \right).$$

The energy dispersion $\varepsilon_J$ of the minibands and the pertaining group velocity $v_{z,J}$ of the carriers entering the above equations are adopted from the tight-binding solution of an atomistic model of the QD stack.

To be able to make statements of practical relevance, it is necessary to extend our modeling by some more empirical elements: The carrier concentration is extremely important for the functioning of the thermoelectric device, as it determines the position of the equilibrium chemical potential of the electrons with respect to the minibands. In the present case of the InAs/GaAs quantum dots, we assume *n*-type conductivity due to electrons supplied by donors predominantly located in the host material. These donors are characterized by a donor concentration $n_{\mathrm{D}}$ and a charge transfer level $\varepsilon_{\mathrm{D}}$. The latter is assumed to lie 10 meV below the bulk conduction band minimum of GaAs for the present example. Moreover, we need to take into account the bulk conduction band states, as these states may be populated by electrons supplied by the donors, and may thus also contribute to the thermoelectric properties. In the following, values from the experimental literature are used for the effective mass of the conduction band electrons of GaAs and InAs. A calculation of the transport relaxation time for carriers in the bulk band yields the well-known increase of $\tau$ proportional to the square root of the energy measured from the conduction band bottom. While the so-defined model provides us with all the information needed to calculate the electronic part of *ZT*, input for the lattice contribution to $\kappa$ must come from elsewhere. At present, we treat $\kappa_{\mathrm{ph}}$ as a numerical parameter that can be set to any available experimental value. In the following, $\kappa_{\mathrm{ph}} = 0.2$ W/(m K) is used.

Results of our model are shown in Fig. 4. We find that the figure of merit *ZT* is highly sensitive to the donor concentration $n_{\mathrm{D}}$ and shows sharp peaks whenever the chemical potential $\mu$ of the electrons falls into one of the minibands. For a particular geometry, e.g. for $a_z = 1.566$ nm and $d_z = 6.26$ nm shown in the left-hand-side panel of Fig. 4, peak values of *ZT* in one miniband are as high as *ZT* = 2. Further analysis of the factors entering the calculation of *ZT* shows that in this miniband regime the thermal conductivity $\kappa$ is dominated by the lattice contribution $\kappa_{\mathrm{ph}}$. While partially occupied minibands contribute already to the electrical conductivity $\sigma$, their contribution to $\kappa_{\mathrm{el}}$ is small because only little entropy can be carried by excitations around the Fermi energy in such a narrow miniband. For high doping concentrations exceeding $5 \times 10^{25}$ m$^{-3}$, the electron chemical potential rises up to the bulk conduction band. Under these conditions, thermal excitations of the carriers high up into empty states in the conduction band are possible. While $\sigma$ for the case of bulk transport is higher than for miniband transport only, $\kappa_{\mathrm{el}}$ increases more steeply than $\sigma$ with increasing $n_{\mathrm{D}}$. Consequently, the figure-of-merit *ZT* drops at very high donor concentrations, when the conductivity becomes dominated by the bulk conduction band.

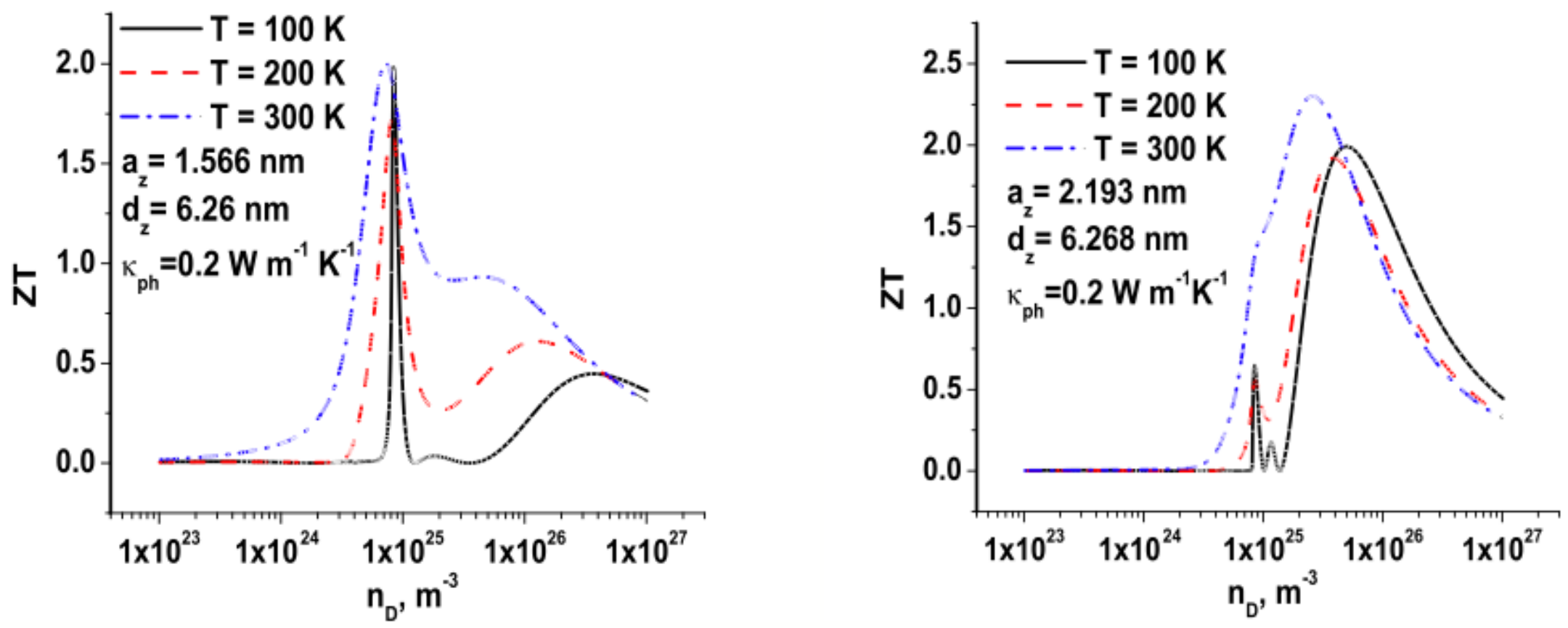

Figure 4. The figure of merit $ZT$ for a 1D stack of InAs/GaAs QDs with height $a_z$=1.566 nm and period $d_z$=6.26 nm (left-hand-side panel) and $a_z$=2.193 nm and $d_z$=6.268 nm (right-hand-side panel) calculated as a function of the concentration of donors $n_D$ with charge transfer level $\varepsilon_D$ at 10 meV below the GaAs conduction band minimum.

Looking at the right-hand-side panel of Fig. 4, one learns that the precise dependence of $ZT$ on the donor concentration may vary from sample to sample, depending on the geometrical parameters of the InAs quantum dots embedded in the GaAs matrix. In the case of somewhat taller QDs ($a_z$ = 2.193 nm) with a stacking period of $d_z$ = 6.268 nm, the figure of merit both at $T$ =200 K and $T$ =300 K shows a single maximum as function of $n_D$. For these geometrical parameters of the QDs, the third miniband in the QD stack comes energetically very close to the onset of the conduction band, and at $T$ =300 K the electrons provided by the donors populate both the highest miniband and the conduction band bottom. Only at low temperatures, $T$ =100 K, small additional peaks at somewhat lower donor concentration $n_D$ due to the minibands are visible. In contrast to pure bulk GaAs, which has a very low figure of merit at room temperature, we predict a significantly enhanced $ZT$ due to nanostructuring. $ZT$ reaches its peak value of 2.3 at $n_D$ = 3 × $10^{25}$ $m^{-3}$ at room temperature, due to added contributions from both the minibands and the bulk conduction band, and again drops off for higher donor concentrations, when the conductivity becomes dominated by bulk states.
A more systematic study of the effect of QD geometry on the thermoelectric figure of merit can be found elsewhere [13]. Already from the examples given here, it is clear that QD crystals hold great promise as materials for thermoelectric converters. However, the results also show that a fine-tuning of the geometrical specifications of the QD inclusions and the suitable donor concentration in the samples is required to exploit the potential of these materials. This is an area of research where fabrication and characterization of the nanostructures should be carried out in close collaboration with simulations of their transport properties to obtain samples with optimum performance.

## 5. Summary and Conclusions

Modeling of materials for optimizing their electrical transport properties must start from a quantum-mechanical modeling step, using density functional theory, or, for very large systems, a tight-binding Hamiltonian to describe the electronic structure of the material. For some properties, e.g. band gaps or magnetic moments, employing the cluster expansion technique allows for a direct search for the optimum in a large configuration space. If dissipative processes play a role for the quantity of interest, as is the case for the thermoelectric figure of merit *ZT*, a hierarchical multi-scale approach is recommended. It comprises solving a Boltzmann transport equation with parameters (band dispersion, transport relaxation time) calculated from a microscopic theory.

## Acknowledgements

Financial support from Deutsche Forschungsgemeinschaft within SPP 1386 'Nanostructured Thermoelectrics' and SFB491 'Magnetic Heterostructures: Spin structure and spin transport' is gratefully acknowledged. V.M.F. has been supported by the European Science Foundation through Exchange Grant No. 2157 within the activity `Arrays of Quantum Dots and Josephson Junctions' and by the German Academic Exchange Service (DAAD). We acknowledge fruitful collaboration with O. G. Schmidt and A. Rastelli on quantum dots arrays for thermoelectrics.